\documentclass[a4paper,10pt,twocolumn]{article}
\usepackage{fixltx2e} % fixes 2-col figure numbering & provides \textsubscript
\usepackage[cm]{fullpage} % Better margins for A4 paper
\usepackage{graphicx}
\usepackage{amsmath}
\usepackage{amssymb}
\usepackage{setspace}
\usepackage{lineno}
\usepackage{authblk}

\usepackage{ucs}               % This and inputenc let us use proper umlaut
\usepackage[utf8x]{inputenc}   % letters without ugly LaTeX commands.

\usepackage{multirow}      % For multi-row table cells
\usepackage{rotating}      % For sideways table labels

\usepackage{ifthen}
\usepackage{color}            % So can use colour (needed for defining cols)

\usepackage{journalabbrevs}

\usepackage[colorlinks,citecolor=blue,linkcolor=blue,urlcolor=blue]{hyperref}
\usepackage{doi}  % makes DOIs in bibliography into links (plainnat only)
\usepackage[authoryear,comma,round]{natbib}      % Bibliography style
\makeatletter
\date{\normalsize Corresponding author: Philip E. Bett (philip.bett@metoffice.gov.uk)}\let\Date\@date
\makeatother

\definecolor{DarkGreen}{rgb}{0.20,0.60,0.20}
\definecolor{orange}{rgb}{1.0,0.50,0.00}

\usepackage[nofancy]{svninfo} % Gives access to SVN info (without fancyhdr)
\svnInfo $Id: cssp_yangtzeverif_forarxiv.tex 7149 2017-08-24 14:37:10Z pbett $ 
\title{\LARGE Seasonal forecasts of the summer 2016 Yangtze River basin rainfall}

\author[1]{\large Philip E.\ Bett}
\author[1,2]{\large Adam A.\ Scaife}
\author[3]{\large Chaofan Li}
\author[1]{\large Chris Hewitt}
\author[1]{\large Nicola Golding}
\author[4]{\large Peiqun Zhang}
\author[1]{\large Nick Dunstone}
\author[1]{\large Doug M.\ Smith}
\author[1]{\large Hazel E.\ Thornton}
\author[5]{\large Riyu Lu}
\author[4]{\large Hong-Li Ren} % He explicitly said to write it like this

\affil[1]{\normalsize Met Office Hadley Centre, FitzRoy Road, Exeter, EX1~3PB, UK}
\affil[2]{\normalsize College of Engineering, Mathematics and Physical Sciences, University of Exeter, Exeter, Devon, UK}
\affil[3]{\normalsize Center for Monsoon System Research, Institute of Atmospheric Physics, Chinese Academy of Sciences, Beijing, China}
\affil[4]{\normalsize Laboratory for Climate Studies, National Climate Center, China Meteorological Administration, Beijing, China}
\affil[5]{\normalsize State Key Laboratory of Numerical Modeling for Atmospheric Sciences and Geophysical Fluid Dynamics, Institute of Atmospheric Physics, Chinese Academy of Sciences, Beijing 100029, China}

\begin{document}
%\linenumbers
\maketitle 
\today 

%\pagebreak

%----------------------------------------------------------------------------
\section*{Abstract}
The Yangtze River has been subject to heavy flooding throughout history, and in recent times severe floods such as those in 1998 have resulted in 
heavy loss of life and livelihoods. Dams along the river help to manage flood waters, and are important sources of electricity for the region. Being able to forecast high-impact events at long lead times therefore has enormous potential benefit. Recent improvements in seasonal forecasting mean that dynamical climate models can start to be used directly for operational services.  The teleconnection from El Niño to Yangtze River basin rainfall meant that the strong El Niño in winter 2015/2016 provided a valuable opportunity to test the application of a dynamical forecast system. 

This paper therefore presents a case study of a \emph{real time} seasonal forecast for the Yangtze River basin, building on previous work demonstrating the retrospective skill of such a forecast.  A simple forecasting methodology is presented, in which the forecast probabilities are derived from the historical relationship between hindcast and observations.  Its performance for 2016 is discussed. The heavy rainfall in the May--June--July period was correctly forecast well in advance. August saw anomalously low rainfall, and the forecasts for the June--July--August period correctly showed closer to average levels.  The forecasts contributed to the confidence of decision-makers across the Yangtze River basin. Trials of climate services such as this help to promote appropriate use of seasonal forecasts, and highlight areas for future improvements.

\noindent This document is \copyright{} Crown Copyright \the\year{} Met Office

\noindent\textbf{Key words}: Seasonal forecasting, flood forecasting, Yangtze basin rainfall, ENSO,  hydroelectricity

%-------------------------------------------------------------------------

\section{Introduction}
The Yangtze River basin cuts across central China, providing water, hydroelectricity and agricultural land for millions of people. The Yangtze has been subject to flooding throughout history \citep[e.g.][]{Plate2002Flood, Yu2009Analysis}, linked to variations in the East Asian monsoon that are sometimes driven by factors such as the El Niño--Southern Oscillation (ENSO; e.g. \citealt{Zhang2016New, Zhang2016Unraveling}).  Large hydroelectric dams along the river and its tributaries, such as the Three Gorges Dam \citep{Meiyan2013threegorges}, have flood defence as their primary responsibility. However, by lowering the water level behind the dam to protect against flooding, less electricity will be produced. There are therefore clear benefits of forecasting impactful rainfall events at long lead times, allowing mitigation planning for flooding and electricity production.

The relationship between ENSO and the East Asian monsoon is complex and not fully understood. However, it has long been clear that a strong El Niño peaking in winter is likely to be followed by above-average rainfall in China the following summer \citep[e.g.][]{He2016East, Xie2016Indowestern, Stuecker2015Combination, Zhang2016New, Zhang2016Unraveling}.  The extreme El Niño event of 1997/1998 was followed by devastating floods in the Yangtze River basin \citep{Zong20001998, Ye20051998, Yuan20172016}: thousands of people died, millions of people were made homeless, and the economic losses ran into billions of CN\yen.  In the subsequent years, much work has gone into better water management and flood prevention, and into improving both the accuracy and communication of climate forecasts, to prevent such a disaster happening again.

Statistical relationships between large-scale climate phenomena and precipitation at more local scales have long been used to produce seasonal forecasts across China, and for the Yangtze in particular \citep[e.g.][]{Zhu2008Statistical, Kwon2009Seasonal, Zou2010East,  ZongJian2011Experiment, Ying2012Improve, Tung2013Improving, Wang2013Subtropical, Peng2014Seasonal, Li2015Improving, Wu2016Seasonal, Xing2016LongLead, Zhang2016Unraveling}.

Recent advances in the dynamical seasonal forecast system developed at the UK Met Office, GloSea5 \citep{MacLachlan2015Global}, have resulted in the development of operational and prototype climate services for the UK in many sectors  \citep[e.g.][]{Svensson2015Longrange, Palin2016Skilful, Clark2017Skilful}. Recent work has shown that GloSea5 also has useful levels of skill for various processes in China \citep{Lu2017Skillful, Bett2017SkillDRAFT}, including for summer precipitation over the Yangtze River basin \citep{Li2016Skillful}, without having to use statistical models based on larger-scale drivers.

In parallel to these findings, \cite{Golding2017Improving} demonstrated that there was a clear demand from users for improved seasonal forecasts for the Yangtze, both from the flood risk and hydropower production communities.  The very strong El Niño that developed during the winter of 2015--2016 \citep{Zhai2016Strong} provided a perfect opportunity to develop a trial operational seasonal forecast using GloSea5 for the subsequent summer of 2016.

We focused on forecasting the mean precipitation for the June--July--August (JJA) period, as that is where \cite{Li2016Skillful}  demonstrated  skill.  However, we also  produced forecasts for the upcoming 3-month period each week, from February to the end of July 2016.  In the last week of each month, a forecast was issued by the Met Office to the Chinese Meteorological Administration (CMA).

In this paper, we describe the observed rainfall in the Yangtze region in summer 2016, and assess how the real time forecasts performed. We describe the data sets used in section~\ref{s:datasets}, and our forecast production methodology in section~\ref{s:forecasts}.  In section~\ref{s:results} we compare the forecasts to the observed behaviour, and discuss possible future developments in section~\ref{s:concs}.

%------------------------------------------------------------------------------

%------------------------------------------------------------------------------
\section{Data sets}\label{s:datasets}
The current operational version of GloSea5 \citep{MacLachlan2015Global} is based on the \textit{Global Coupled 2} (GC2) configuration  of the HadGEM3 global climate model, described in detail in \cite{Williams2015The} and references therein. Within HadGEM3-GC2, the atmospheric component (the Met Office Unified Model, UM, \citealt{Walters2017Met})  is coupled to the JULES land surface model \citep{Best2011Joint}, the NEMO ocean model \citep{Megann2014GO50, Madec2008nemo} and the CICE sea ice model \citep{Rae2015Development, Hunke2010cice}.  The atmosphere is modelled on a grid of 0.83° in longitude and 0.55° in latitude with 85 levels vertically, including a well-resolved stratosphere; the ocean model is modelled on a  0.25° horizontal grid with 75 levels vertically.

Using this configuration, GloSea5 runs operationally, producing both forecasts and corresponding hindcasts  (used to bias-correct the forecasts).  Each day, two initialised forecasts are produced, running out to 7 months.  To produce a complete forecast ensemble for a given date, the last three weeks of individual forecasts are collected together to form a  42-member lagged forecast ensemble.  At the same time, an ensemble of hindcasts is produced each week. As described by \cite{MacLachlan2015Global}, 3 members are run from each of 4 fixed initialisation dates in each month, for each of the 14 years covering 1996--2009.  The full hindcast ensemble is made by collecting together the four hindcast dates  nearest to the forecast initialisation date, yielding a 12-member, 14-year hindcast.  Note that the hindcast was extended in May 2016 to cover 23 years (1993--2015). 

This operational hindcast is not intended to be used for skill assessments: with only 12 members, skill estimates would be biased low \citep{Scaife2014Skillful}. However, a separate, dedicated  hindcast was produced for skill assessment, with 24 members and 20 years.  Using that hindcast, we find a correlation skill of 0.56 for summer Yangtze rainfall, statistically indistinguishable from the previous value of 0.55 found by \cite{Li2016Skillful}.

We use precipitation data from the Global Precipitation Climatology Project (GPCP) as our observational data set. This is derived from both satellite data and surface rain gauges, covering the period from 1979 to the present at 2.5° spatial resolution \citep{Adler2003Version2}. The verification we present here uses version 2.3 of the data \citep{Adler2016New}. Only version 2.2 was available when we started our operational trial, although we have confirmed that the choice of v2.2 or v2.3 makes negligible difference to our forecasts or results.

%-------------------------------------------------------------------------
\section{Forecast production}\label{s:forecasts}
Typically when producing a seasonal forecast, the distribution of forecast ensemble members is used to represent the forecast probability distribution directly. However, experience has shown that the GloSea5 ensemble members may contain anomalously small signals, such that the predictable signal only emerges through averaging a large ensemble \citep{Scaife2014Skillful, Eade2014Do}. While this effect is less pronounced in subtropical regions like the Yangtze Basin, it is still present \citep{Li2016Skillful}. 

We therefore implemented a simple precipitation forecasting methodology, based entirely on the historical relationship between the hindcast ensemble means and the observed precipitation, averaged over the Yangtze River basin region (91°--122°E,  25°--35°N, following \citealt{Li2016Skillful}), for the season in question.  The prediction intervals, derived from the linear regression of the hindcasts to the observations \citep[e.g.][]{Wilks2011Statistical}, provide a calibrated forecast probability distribution.  

\begin{figure}
\centering
\includegraphics[width=0.5\textwidth]{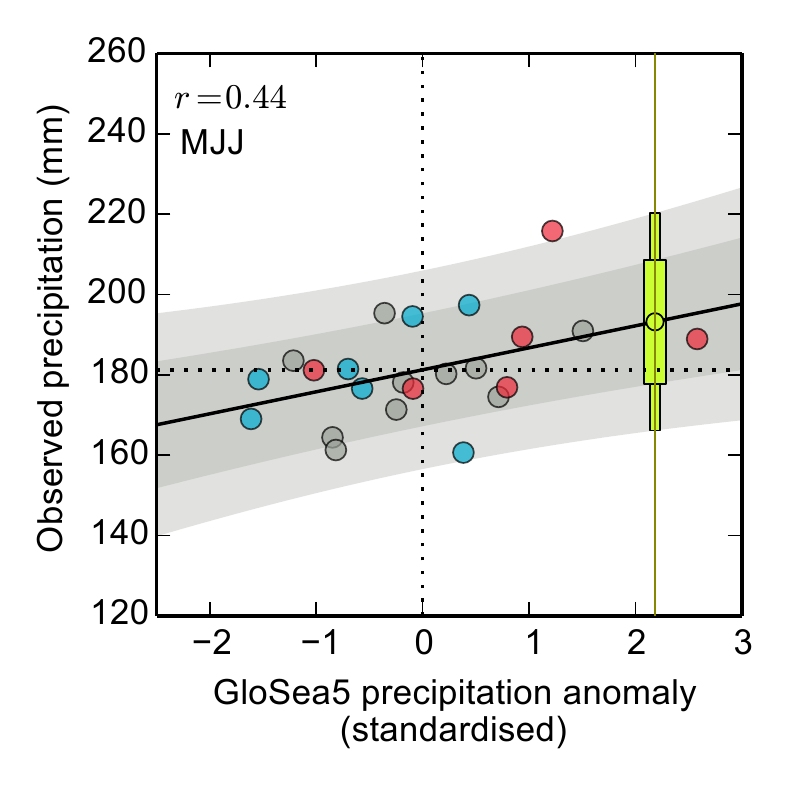}
\includegraphics[width=0.5\textwidth]{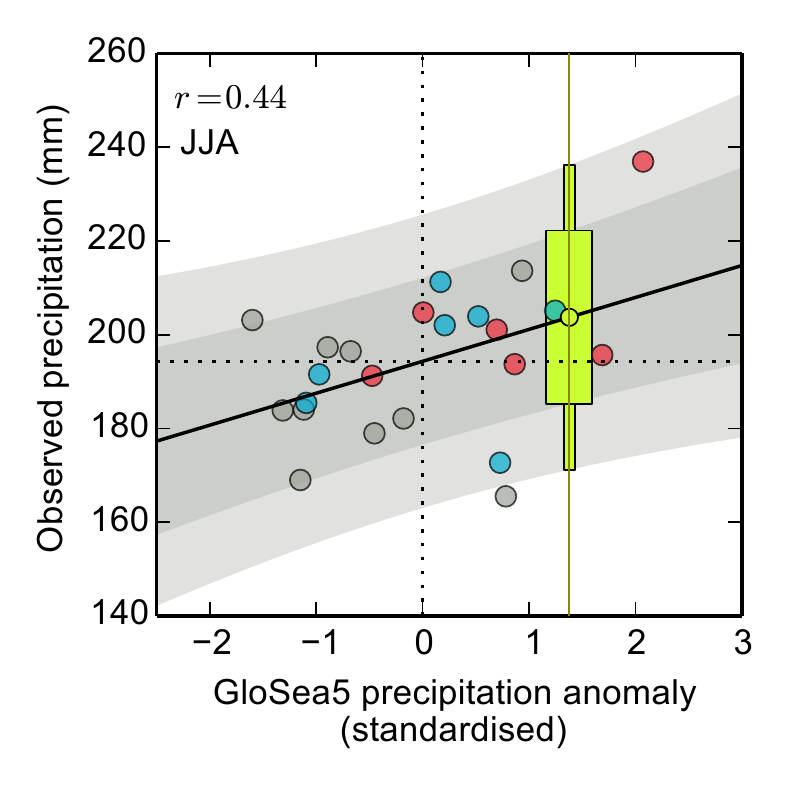}
\caption{The forecast for MJJ (top, produced on 25th April 2016) and JJA (bottom, produced 23rd May 2016), using the  GPCP  observations.   Observation/hindcast points are colour-coded according to their observed winter ENSO index: red points are El Niño years, blue points are La Niña years, and grey points are neutral. The horizontal width of the green forecast bars is the standard error on the ensemble mean, i.e. the forecast ensemble spread divided by the number of ensemble members.   The correlation $r$ between hindcast and observations is marked on each panel (co-incidentally the same when rounded). }
\label{f:fcissued}
\end{figure}

This is illustrated in Figure~\ref{f:fcissued}, where we show the precipitation forecasts issued in late April for MJJ, and in late May for JJA.  The distribution of hindcasts and observations is shown as a scatter plot, with the ensemble mean forecast also included as a green circle. The uncertainty in the linear regression (grey) determines the forecast probabilities (green bars).  The GloSea5 data is shown in standardised units, that is, the anomaly of each year from the mean, as a fraction of the standard deviation of hindcast ensemble means.  The observations on the vertical axis are presented as seasonal means of monthly precipitation totals. The relationship with ENSO is indicated though colour-coding of the hindcast points: Years are labelled as El Niño (red) or La Niña (blue) according to whether their Oceanic Niño Index,\footnote{\url{http://www.cpc.noaa.gov/products/analysis_monitoring/ensostuff/ensoyears.shtml}.} based on observed sea surface temperature anomalies in the Niño 3.4 region, is above $0.5\,\mathrm{K}$ or below $-0.5\,\mathrm{K}$,  respectively.

Forecasts like those shown in Figure~\ref{f:fcissued} are produced each week, using the forecast model runs initialised each day of the preceding 3 weeks to generate the 42-member ensemble. We only issued the forecast produced near the end of each month. The full release schedule is described in Table~\ref{t:releases}.

\begin{table} %[p]
\caption{\label{t:releases}Dates of the monthly forecast releases.}
\begin{tabular}{lll}\hline
Release & Forecast date & Forecast season \\\hline\hline
1 & 22nd Feb 2016 & MAM \\ 
2 & 21st Mar 2016 & AMJ \\ 
3 & 25th Apr 2016 & MJJ \\ 
4 & 23rd May 2016 & JJA \\ 
5 & 27th Jun 2016 & JAS \\ 
6 & 25th Jul 2016 & ASO \\\hline
\end{tabular}
\end{table}

It is important to note that, due to the linear regression method we employ, our forecast probabilities are explicitly linked to both the hindcasts and the observations.  The correlations between hindcasts and observations are biased low due to the smaller size of the hindcast ensemble compared to the forecast ensemble -- a larger hindcast ensemble would not necessarily alter the gradient of the linear regression, but would reduce its uncertainty. Our forecast probabilities are therefore conservative (likely to be too small).

The forecast information provided was designed to show very clearly and explicitly the uncertainties in the forecast system, to prevent over-confidence on the part of potential decision-makers.  In addition to the scatter plot showing the forecast and the historical relationship (Figure~\ref{f:fcissued}), we also provided the probability of above-average precipitation as a `headline message'. This was accompanied by a contingency table showing the hit rate and false alarm rate for above-average forecasts over the hindcast period. For the MJJ and JJA forecasts, these are shown in Tables~\ref{t:contingencyMJJ} and~\ref{t:contingencyJJA}.

\begin{table} %[p]
\caption{\label{t:contingencyMJJ} The contingency table for forecasts of above-average precipitation for the Yangtze region in MJJ, produced on 25th April 2016.  The event counts are based on the GPCP observations and ensemble mean hindcasts shown in Figure~\ref{f:fcissued}.  The hit rate is the ratio of the number of hits to the number of times above-average conditions were observed. The false alarm rate is the ratio of the number of false alarms to the total number of observed below-average years.}
\begin{tabular}{cc|cc} \hline\hline
\multicolumn{2}{c|}{Above-average} & \multicolumn{2}{c}{Observed} \\
\multicolumn{2}{c|}{precipitation} & Yes & No \\ \hline
\multirow{4}{*}{\begin{sideways}Predicted\end{sideways}}
 & \multirow{2}{*}{Yes} & 6      & 4 \\ 
 &                      & Hits   & False alarms \\ %\cline{2-4}
 & \multirow{2}{*}{No}  & 4      & 9 \\ 
 &                      & Misses & Correct rejections \\ \hline\hline
\multicolumn{2}{l}{Hit rate:}         &  \multicolumn{2}{c}{60\%} \\
\multicolumn{2}{l}{False alarm rate:} &  \multicolumn{2}{c}{30\%} \\ \hline\hline
\end{tabular}
\end{table}

\begin{table} %[p]
\caption{\label{t:contingencyJJA} The contingency table for forecasts of above-average precipitation in JJA,  produced on 23rd May 2016, similar to Table~\ref{t:contingencyMJJ}.}
\begin{tabular}{cc|cc} \hline\hline
\multicolumn{2}{c|}{Above-average} & \multicolumn{2}{c}{Observed} \\
\multicolumn{2}{c|}{precipitation} & Yes & No \\ \hline
\multirow{4}{*}{\begin{sideways}Predicted\end{sideways}}
 & \multirow{2}{*}{Yes} & 9      & 3 \\ 
 &                      & Hits   & False alarms \\ %\cline{2-4}
 & \multirow{2}{*}{No}  & 3      & 8 \\ 
 &                      & Misses & Correct rejections \\ \hline\hline
\multicolumn{2}{l}{Hit Rate:}         &  \multicolumn{2}{c}{75\%} \\
\multicolumn{2}{l}{False Alarm Rate:} &  \multicolumn{2}{c}{25\%} \\ \hline\hline
\end{tabular}
\end{table}

%------------------------------------------------------------------------------

\section{Results}\label{s:results}
The observed precipitation in May, June, July and August 2016 is shown in Figure~\ref{f:fcmapsobs}.  We use standardised units here to show the precipitation anomaly relative to the historical variability over the hindcast period (1993--2015). It is clear that the most anomalously high rainfall was in May and June, and largely in the eastern half of the basin. July was close to normal overall when considering the box we were forecasting for, although there were disastrous floods further north. August had anomalously low rainfall across most of the region.  \cite{Yuan20172016} have examined the observed  summer 2016 rainfall in China and the Yangtze River basin in detail, including its relationship to larger-scale drivers: the anomalously low rainfall in August 2016 is in marked contrast to the situation in 1998, and is related to the behaviour of Indian Ocean temperatures and the Madden--Julian Oscillation during the summer.

\newcommand{\mapfigw}{0.48\textwidth}

\begin{figure*}[h]
\centering
\includegraphics[width=\mapfigw]{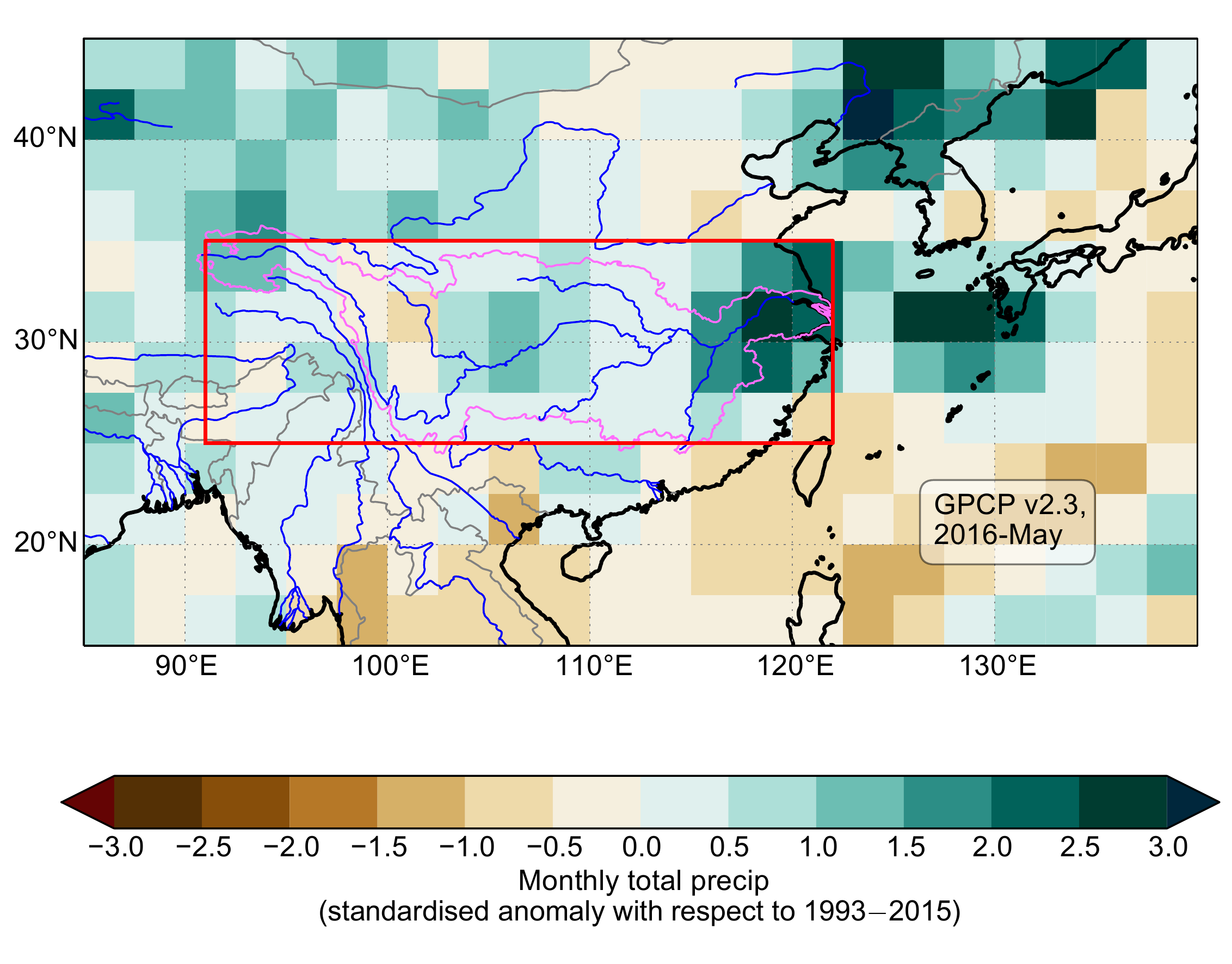}
\includegraphics[width=\mapfigw]{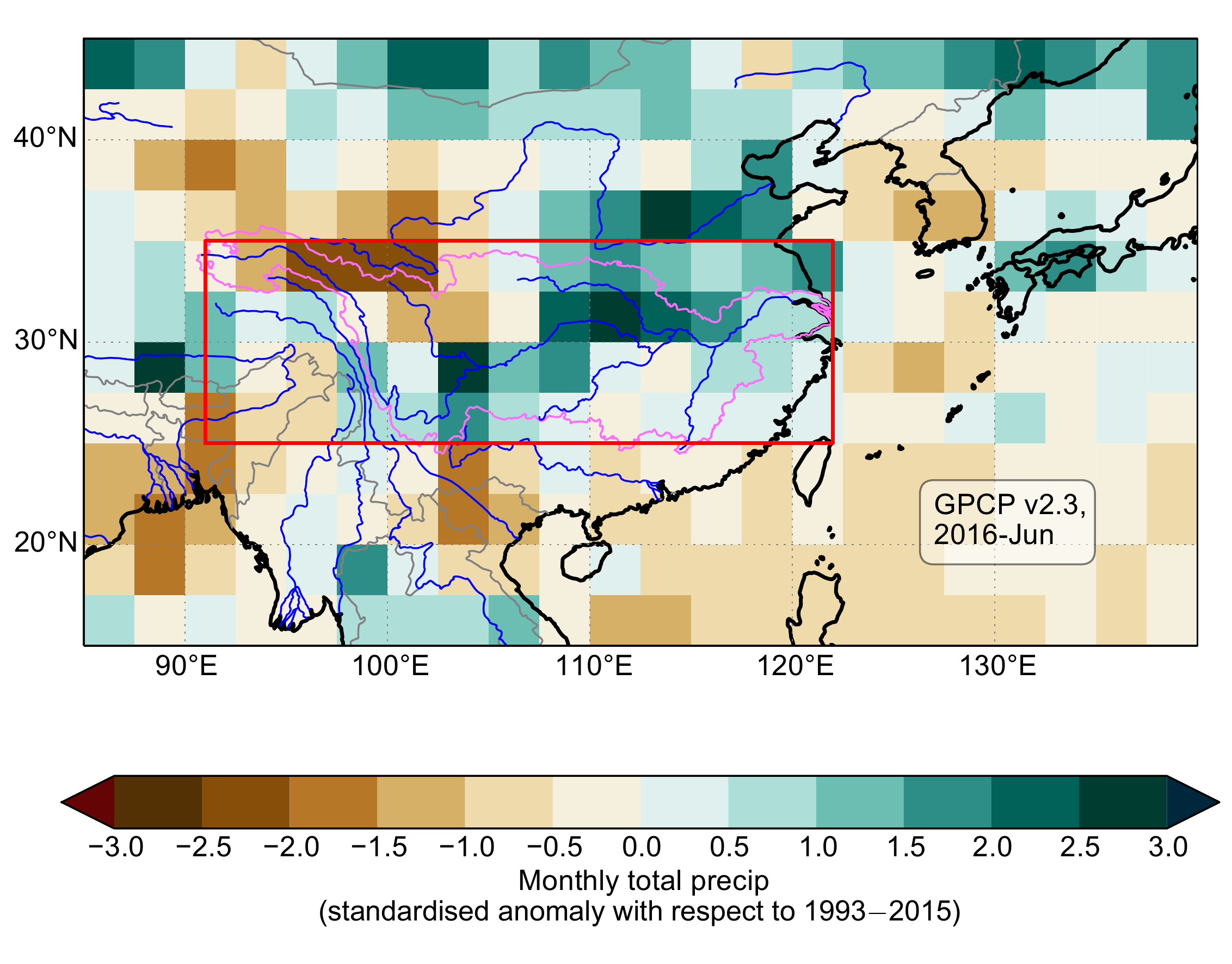}\\
\includegraphics[width=\mapfigw]{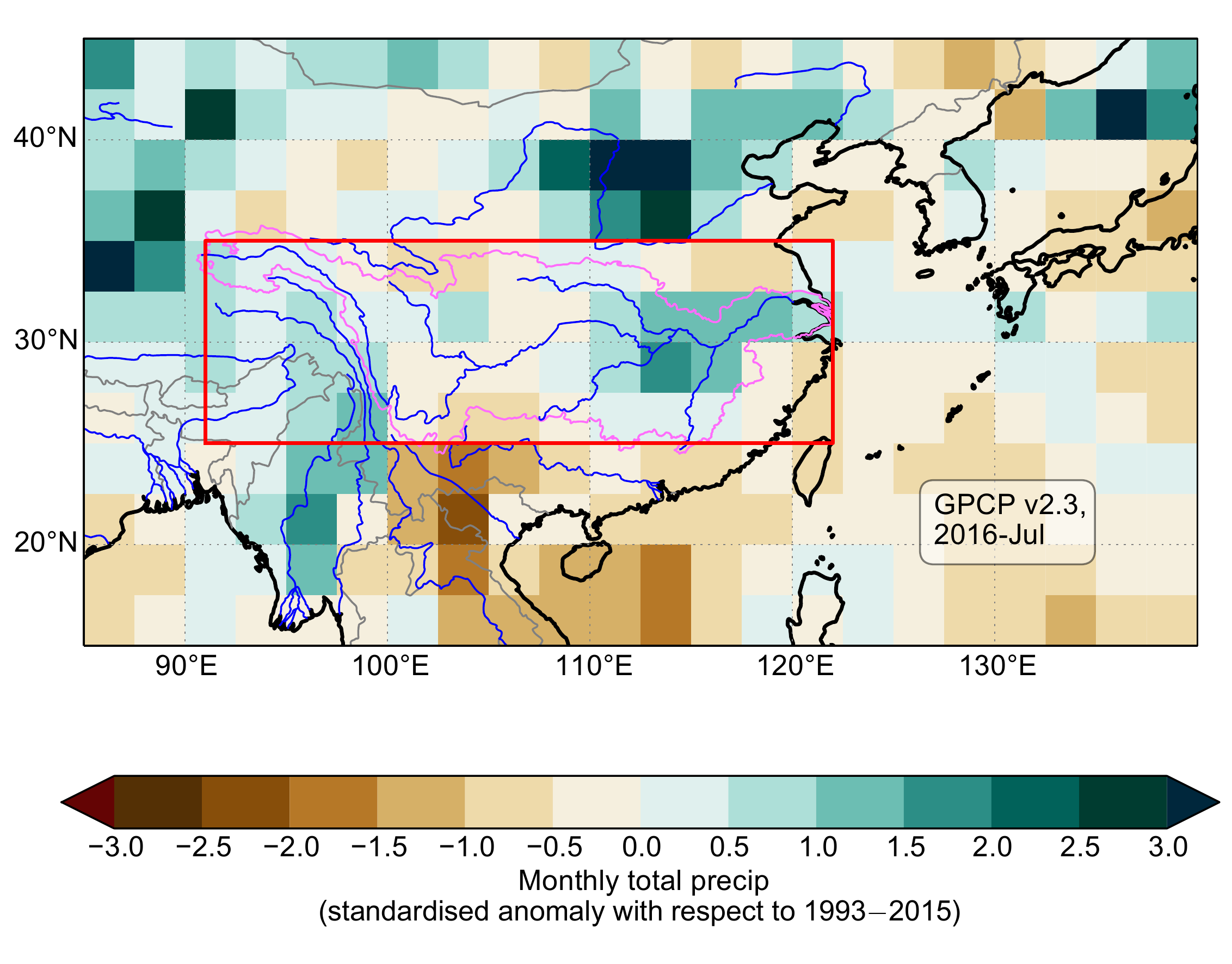}
\includegraphics[width=\mapfigw]{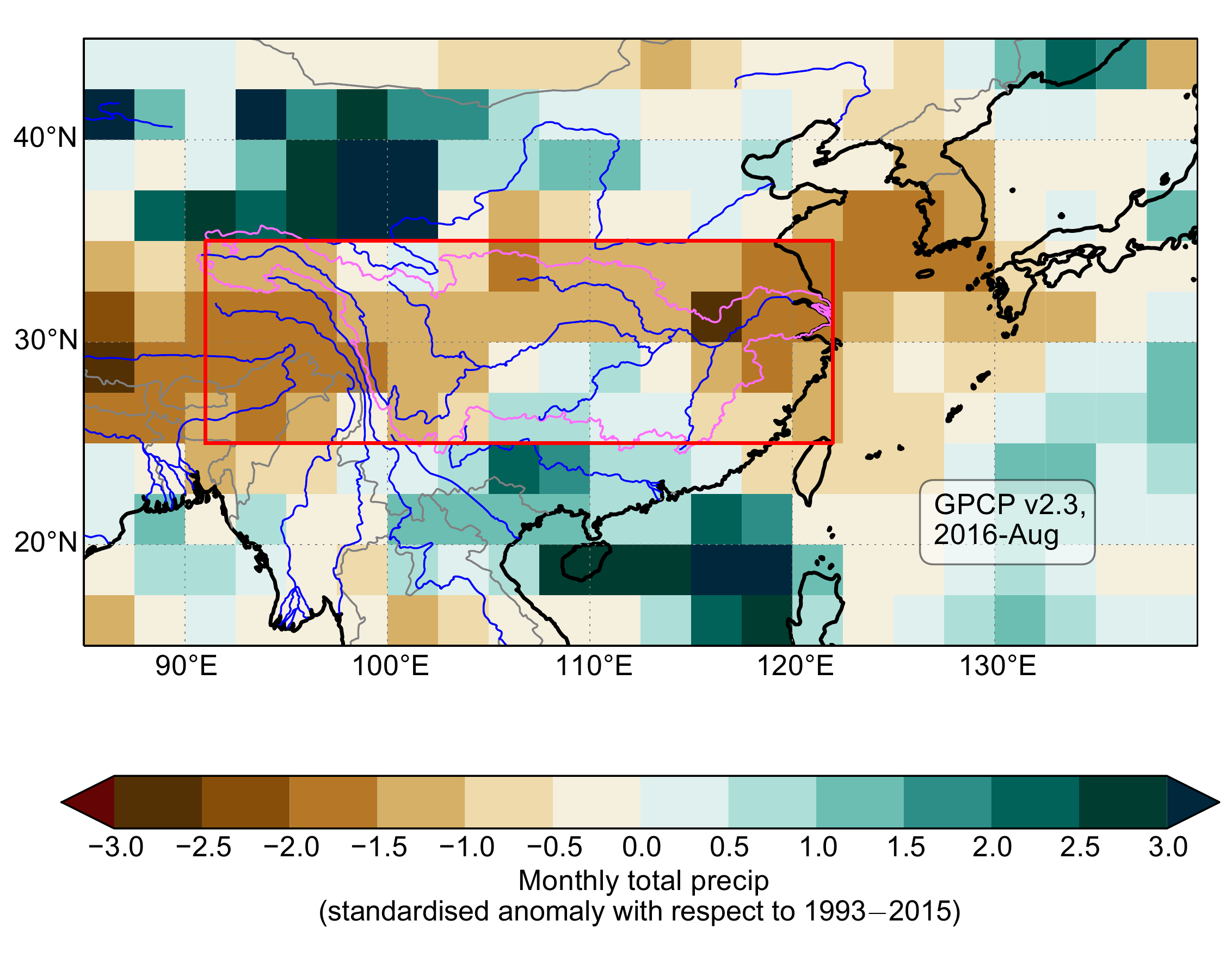}
\caption{Observed precipitation from GPCP v2.3 for May, June, July, and August (as labelled), in standardised units with respect to the 1993--2015 period.  The Yangtze box used for the forecasts is marked as a red rectangle, with a pink polygon showing the physical Yangtze River catchment.  Major rivers are marked in blue.  }
\label{f:fcmapsobs}
\end{figure*}

\begin{figure}[h]
\centering
\includegraphics[width=0.5\textwidth]{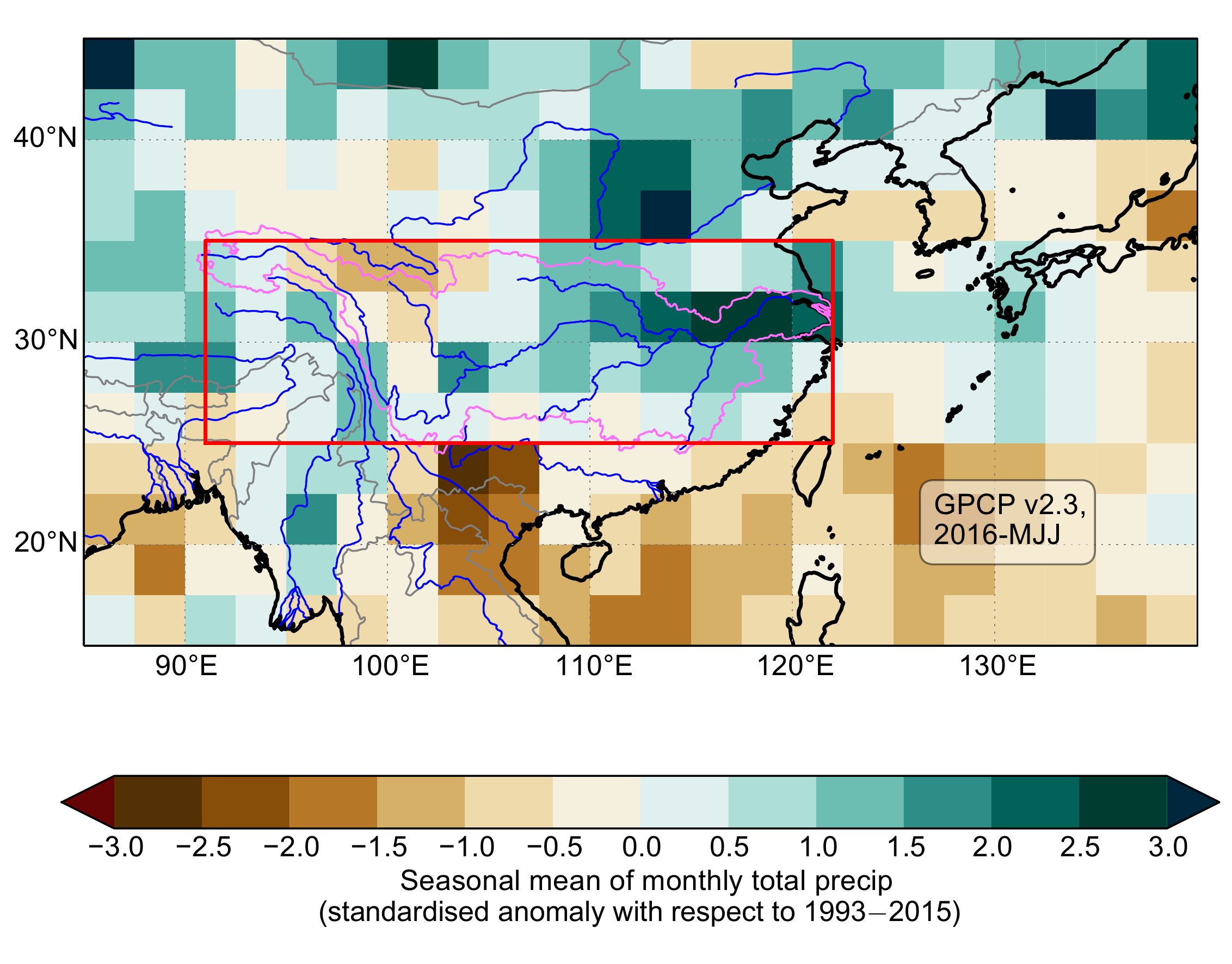}
\includegraphics[width=0.5\textwidth]{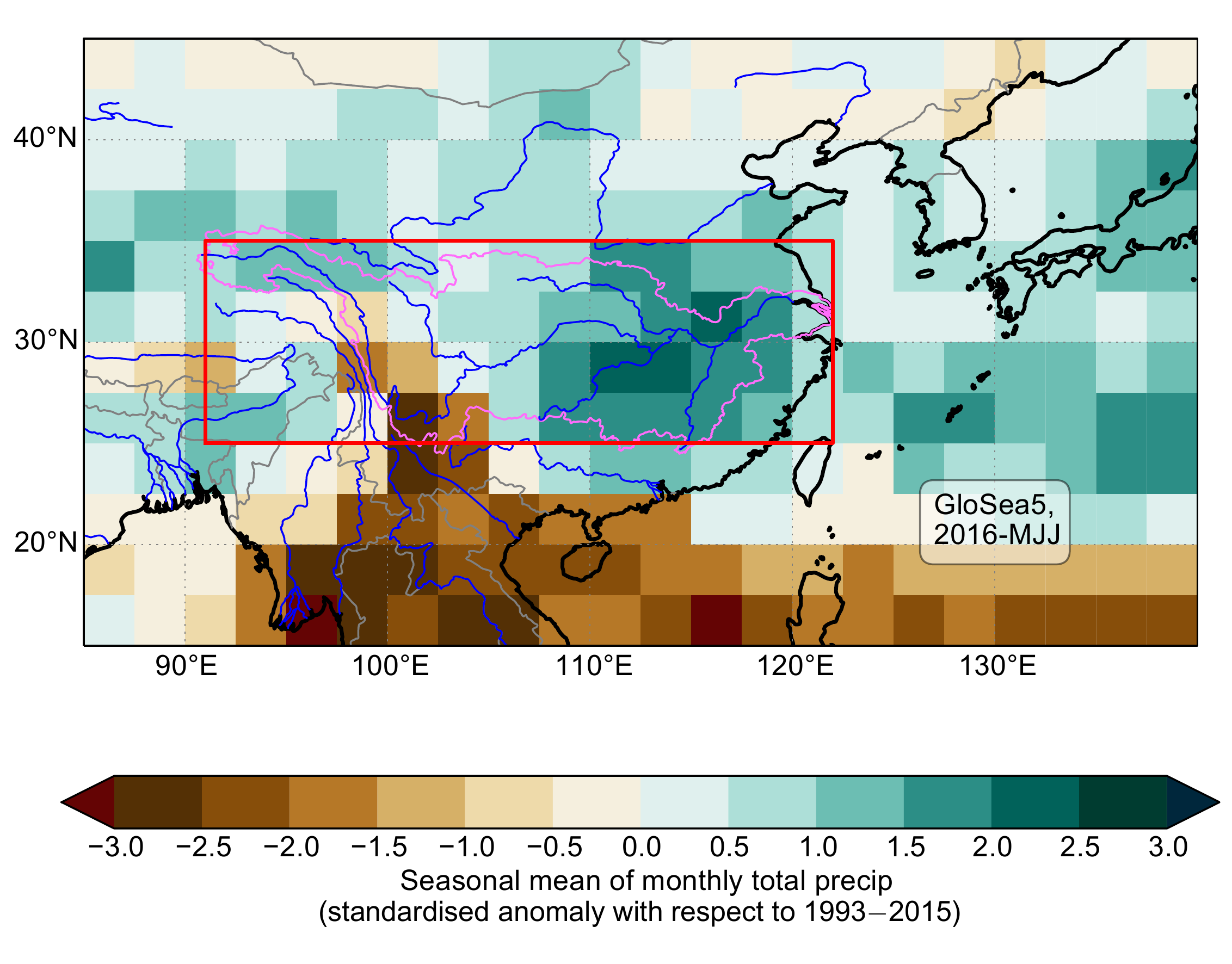}
\caption{Mean precipitation for 2016-MJJ  in the GPCP observations (top), and the forecast signal (bottom), in standardised units.  The GloSea5 data has been regridded to match the lower-resolution observations.
 }
\label{f:fcmapsmjj}
\end{figure}

\begin{figure}[h]
\centering
\includegraphics[width=0.5\textwidth]{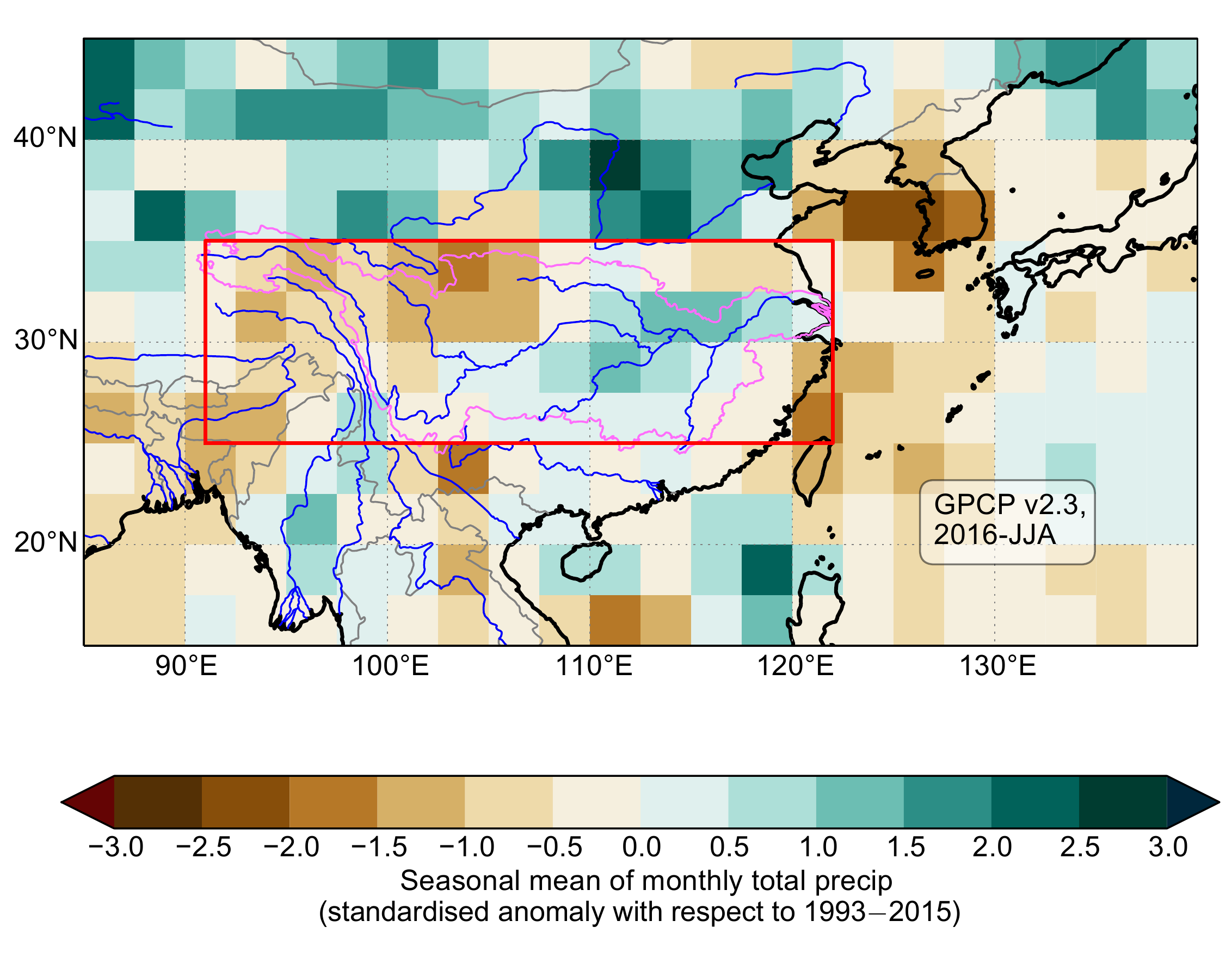}
\includegraphics[width=0.5\textwidth]{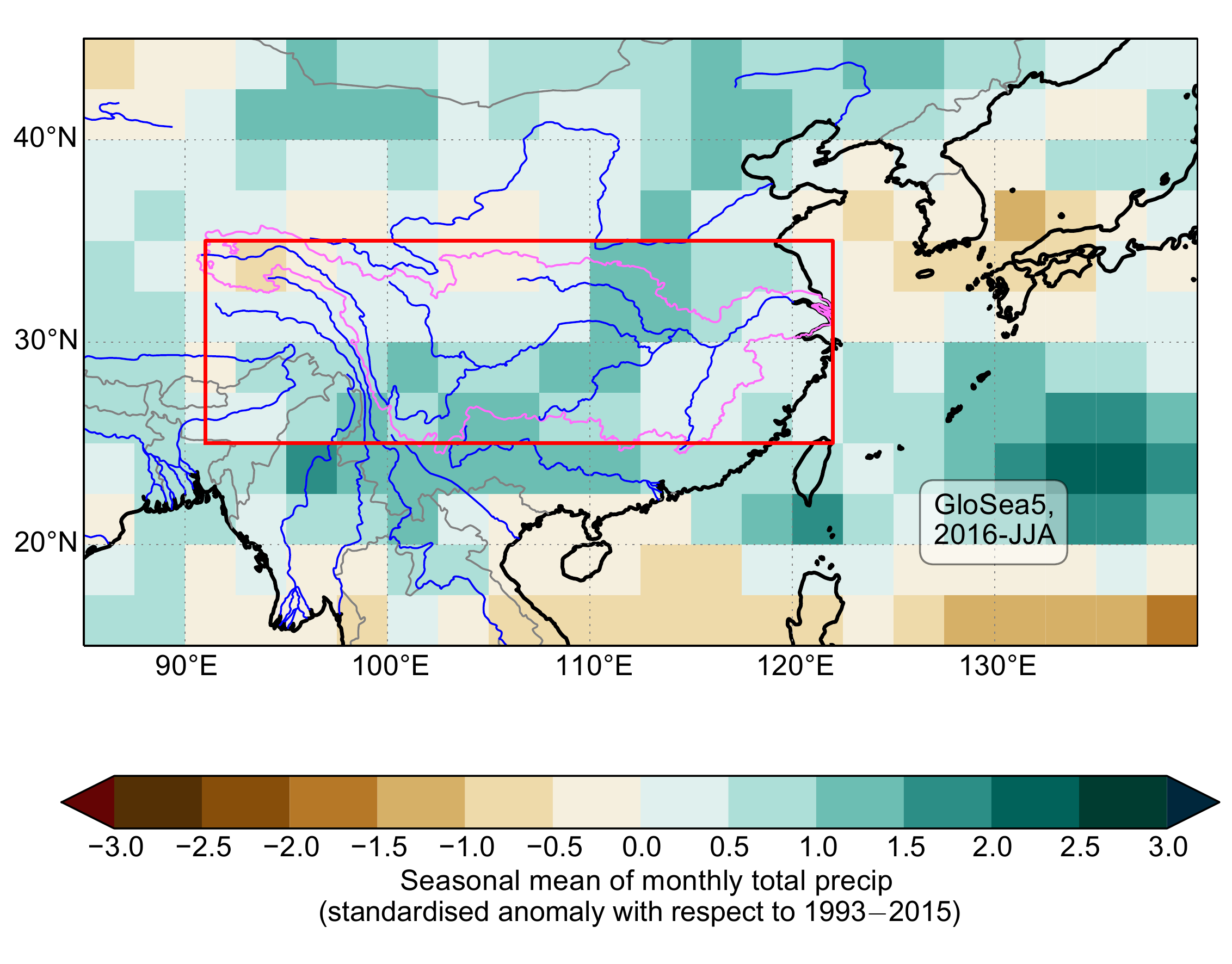}
\caption{Mean precipitation for 2016-JJA in the GPCP observations (top), and the forecast signal (bottom), in standardised units. The GloSea5 data has been regridded to match the lower-resolution observations.
}
\label{f:fcmapsjja}
\end{figure}

Figures~\ref{f:fcmapsmjj} and~\ref{f:fcmapsjja} show the 3-month mean precipitation anomalies for MJJ and JJA respectively, for both GPCP and the forecast averages from the GloSea5 model output. While we do not expect the spatial patterns to match in detail (considering the skill maps of \citealt{Li2016Skillful}), the overall signal is similar to the observations, with stronger anomalous precipitation in the eastern region in MJJ, and closer-to-average precipitation in JJA.

\begin{figure*} %0.8
\centering\includegraphics[width=0.66\textwidth]{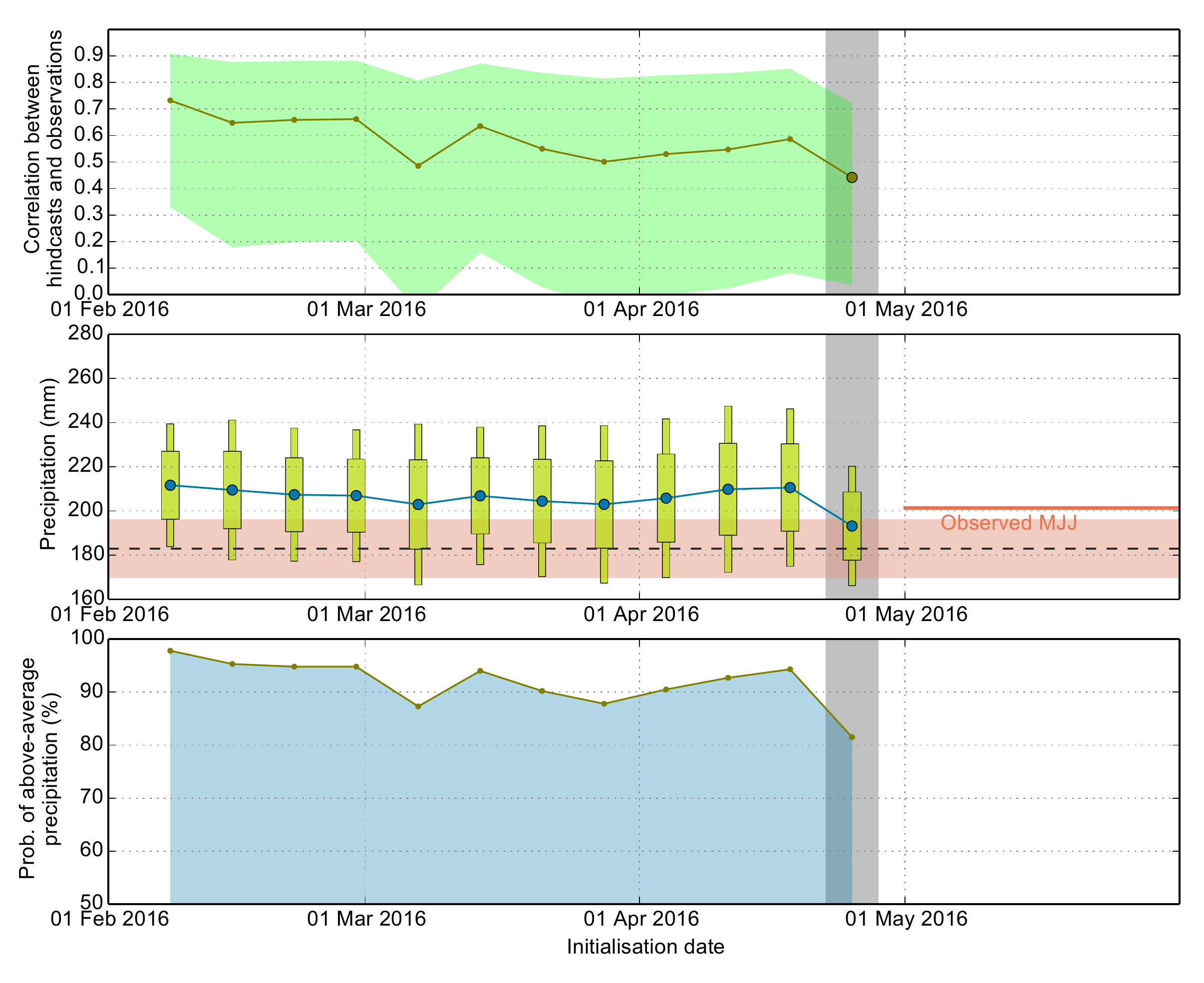}
\caption{Time series showing the behaviour of the MJJ forecasts and hindcasts with lead time. Top: Correlation between  observations and the operational hindcasts available each week. The final point was produced using 23 years, whereas only 14 were available before that. The shading indicates 95\% confidence intervals using the Fisher Z-transformation.  Middle: The forecast signal shown as 95\% and 75\% prediction intervals (boxes) and the ensemble mean (blue line). The observed precipitation is marked as an orange horizontal line from May. The observed historical mean and standard deviation over the hindcast period are marked as a dashed line and orange shading respectively.  Bottom: The forecast probability of above-average precipitation.  The final forecast issued for MJJ, produced on 25th April, is highlighted with a grey vertical bar. }
\label{f:veriftsMJJ}
\end{figure*}

\begin{figure*}
\centering\includegraphics[width=0.66\textwidth]{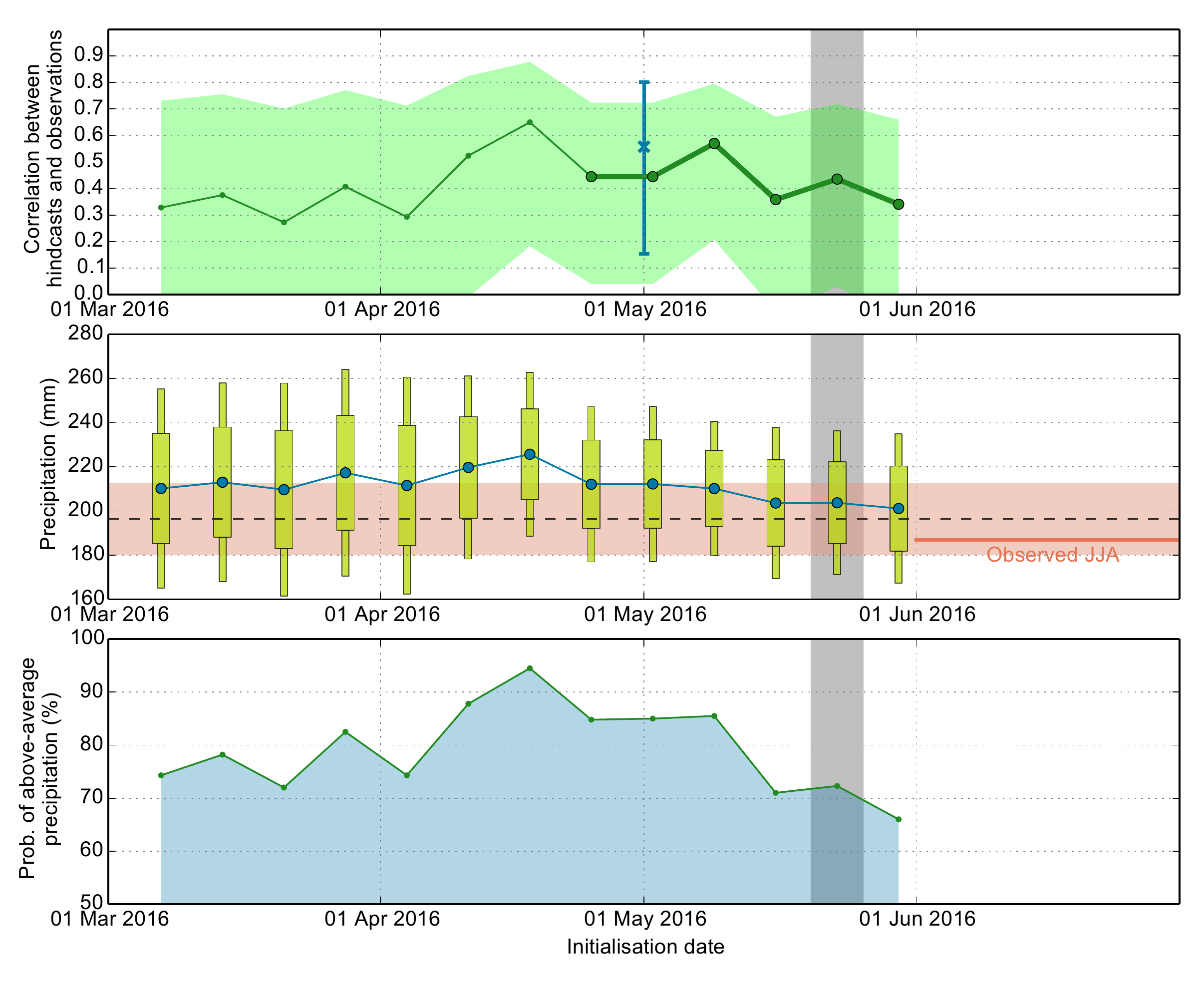}
\caption{Time series showing the behaviour of the JJA forecasts and hindcasts with lead time, following the same format as Figure~\protect\ref{f:veriftsMJJ}.  In the top panel (hindcast--observations correlation), the line becomes thicker when 23 years of hindcast are available. We mark with a blue cross and error bar the correlation skill derived from the assessment hindcast (see text for details).   The final forecast issued for JJA, on 23rd May, is highlighted across all panels.}
\label{f:veriftsJJA}
\end{figure*}

We examine our forecasts for the Yangtze basin box more quantitatively in Figures~\ref{f:veriftsMJJ} and~\ref{f:veriftsJJA}, where we show the variation with lead time of the hindcast--observation correlation, the 2016 forecast signal, and the probability of above-average precipitation, for MJJ and JJA respectively.  Neither the hindcast--observation correlation nor the forecast signal vary significantly with lead time; indeed, they are remarkably consistent back to  3 months before the forecast season, and when the 23-year hindcast is introduced in May.

The forecasts did a good job of giving an indication of precipitation in the coming season. For MJJ, the forecast gave a high probability of above-average precipitation (80\%), and it was observed to be above average. In JJA, the mean precipitation was observed to be slightly below average, due to the strong drier-than-average signal in August, although it was within a standard deviation of the interannual variability. While our forecast marginally favoured wetter than average conditions (65\% probability of above-average rainfall), it was correctly near to the long-term mean, and the observed value was well within the forecast uncertainties.

% ------------------------------------------------------------------------------

% ------------------------------------------------------------------------------
\section{Discussion and conclusions}\label{s:concs}
The heavy rainfall in the Yangtze River region in early summer 2016 was at a similar level to  that of 1998, and caused heavy flooding \citep{WMOstatement2017, Yuan20172016}.  While deaths due to the flooding were roughly an order of magnitude fewer than those caused by the 1998 floods (i.e. hundreds rather than thousands of lives), the economic losses nevertheless ran into tens of billions of CN\yen.  Furthermore, it was reported that insurance claims, mostly from agricultural losses, amounted to less than 2\% of the total economic loss, suggesting significant levels of underinsurance \citep{Podlaha2016July}.  The prior experience of the 1998 El Niño-enhanced flooding, and the high levels of awareness of the strong El Niño in winter 2015/2016, meant that dams along the Yangtze were prepared in anticipation of high levels of rainfall.  Our forecasts from GloSea5, produced using the simple methodology described here,  contributed to the confidence of users adapting to the impending rainfall (Golding et al., in prep).

Our verification has shown that our forecasts gave a good indication of the observed levels of precipitation for both MJJ and JJA averages over the large Yangtze Basin region.  A greater degree of both spatial and temporal resolution -- splitting the basin into upper and lower sections, and producing additional forecasts at a monthly timescale -- would of course be preferable to users.  However, smaller regions and shorter time periods may well be less skillful, so further work is needed to assess how best to achieve skillful forecasts in these cases.

One significant improvement would be to increase the ensemble size of the hindcast. During 2017 the GloSea5 system was changed from 3 hindcast members per start date to 7. This could result in noticeable improvements in forecasts like those described here, as the hindcast--observations relationship will be less uncertain, especially when a predictable signal is present such as from El Niño.

We will be issuing forecasts again in 2017. However, unlike 2016, in 2017 there are no strong drivers such as El Niño.  Nevertheless, understanding the behaviour of the forecast system under such conditions will be informative, for both the users and the producers of the forecasts.  Ultimately, trial climate servives such as this help to drive forecast development, improve understanding of forecast uncertainties, and promote careful use by stakeholders in affected areas.

%--------------------------------------------------------------------------

%----------------------------------------------------------
\section*{Acknowledgements}
This work and its contributors (PB, AS, ND, DS, CH, NG) were supported by the UK--China Research \& Innovation Partnership Fund through the Met Office Climate Science for Service Partnership (CSSP) China as part of the Newton Fund. 
CL and RL were supported by the National Natural Science Foundation of China (Grant No.\ 41320104007).
HR was supported by the Project for Development of Key Techniques in Meteorological Operation Forecasting (YBGJXM201705).
The trial forecast service was first suggested by AS in 2015.

% https://www.esrl.noaa.gov/psd/data/gridded/data.gpcp.html#citations
GPCP precipitation data provided by the NOAA/OAR/ESRL PSD, Boulder, Colorado, USA, from their web site at \url{http://www.esrl.noaa.gov/psd/}. 

\vspace{1cm} % This was needed just to help compilation
% - otherwise, the shapefile url was broken across pages
%   in the first pass before bib/refs.

% http://worldmap.harvard.edu/data/geonode:ch_wtrshed_30mar11
The Yangtze river basin shapefile used in the maps was obtained from \url{http://worldmap.harvard.edu/data/geonode:ch_wtrshed_30mar11} and is based on the watersheds shown in the China Environmental Atlas (2000), \copyright\ Chinese Academy of Science, Environmental Data Center.

%----------------------------------------------------------

%===================================================================
%\clearpage
\bibliographystyle{plainnat}
\bibliography{cssp_yangtzeverif}
%===================================================================

%\clearpage %pagebreak
%\listoftables
%\listoffigures

\end{document}